\documentclass[useAMS,usenatbib]{mn2e}

\usepackage{graphicx}

\title[Genetic algorithm to model protoplanetary discs]{The use of genetic algorithm to model protoplanetary discs}

\author[A. Hetem Jr. and J. Gregorio-Hetem]{Annibal Hetem Jr. $^{1}$\thanks{E-mail:annibal.hetem.jr@usa.net; jane@astro.iag.usp.br} and Jane Gregorio-Hetem$^{2}$
\footnotemark[1]\thanks{Based on observational data obtained in the Pico dos Dias Observatory (LNA/MCT) - Brazil}\\
$^{1}$Funda\c c\~ao Santo Andr\'{e} FAFIL, Av. Pr\'\i ncipe de Galles, 821, Santo Andr\'{e}, SP - Brazil\\
$^{2}$Universidade de S\~ao Paulo IAG-USP, Rua do Mat\~ao, 1226, S\~ao Paulo, SP - Brazil}

\begin{document}


\pagerange{\pageref{firstpage}--\pageref{lastpage}} \pubyear{2007}

\maketitle

\label{firstpage}

\begin{abstract}
The protoplanetary discs of T Tauri and Herbig Ae/Be stars have been studied by
using geometric disc models to fit their spectral energy distribution (SED). The simulations 
provide means to reproduce the signatures of different circumstellar structures,
which are related to different levels of infrared excess. Aiming to improve our previous
model that assumed a simple flat disc configuration, in the present work we adopt a
reprocessing flared disc model that considers hydrostatic, radiative equilibrium \citep{DDN2001}. 
We developed a method to optimise the parameters estimation based on genetic algorithms 
(GA). This paper is dedicated to describe the implementation of the new code,
which has been applied for Herbig stars from the {\it Pico dos Dias Survey} catalogue,
in order to illustrate the quality of the fitting for a variety of SED shapes. The star
AB Aur was used as a test of the GA parameters estimation, demonstrating that the
new code reproduces successfully a canonical example of the flared disc model. The
GA method gives good quality of fittings, but the range of input parameters must
be chosen with caution, since unrealistic disc parameters can be derived. The flared
disc model is confirmed to fit the flattened SEDs typical from Herbig stars, however the
embedded objects (increasing SED slope) and debris discs (steeper SED) are
not well fitted with this configuration. Even considering the limitation of the derived
parameters, the automatic process of SED fitting provides an interesting tool for the
statistical analysis of the circumstellar luminosity of large samples of young stars.
\end{abstract}


\section{Introduction}

\subsection{Background}

The use of a model to study the disc geometry and the circumstellar emission from young stars is 
part of a detailed analysis that our group have developed for a large sample of T Tauri stars (TTs) and Herbig 
Ae/Be stars (HAeBes) detected by the {\it Pico dos Dias Survey - PDS} \citep{PDS1,PDS2,Torres98}. 
In a general context, the interest is to discuss the geometry of protoplanetary discs in terms of disc evolution. 
While the accretion processes occur for most of the low mass stars with age about 1 Myr, the 
disc dissipation is expected to happen within 10 Myr \citep{Haisch01,Bergin04,Forrest04,Dalessio05,Sicilia06}. 

Among different circumstellar structures that can be used to explain the spectral energy distribution 
(SED) from young stars, in a previous work \cite{GHH02} (hereafter GH02) used a flat disc model, which 
re-irradiates the energy absorbed from the star, leading to infrared excess even in the absence of any 
accretion luminosity \citep{Adams87,Kenyon87,Strom93}. 
Assuming that flat disc models are valid when there is no accretion, GH02 adopted the simple model of 
passive disc to study weak-line TTs (WTTs).
The occurrence of accretion is not expected for these objects, as indicated by their low H$\alpha$ 
equivalent width ($W_{{\rm H}\alpha}$).
They selected 27 {\it PDS} stars showing $W_{{\rm H}\alpha} < 10$ \AA{}, a usual criterion to 
distinguish WTTs from classical TTs (CTTs) \citep{Appenz89}.  
The individual contribution of the circumstellar components to the total emitted flux was compared with stellar 
characteristics in order to better classify the sample in agreement with an evolutionary sequence. 
GH02 concluded that almost half of the studied WTTs have 
considerable circumstellar luminosity ($\sim 30$ percent of the total luminosity)
confirming that the absence of accretion does not imply in the lack of circumstellar material, as verified by 
\cite{Andre94} in a comparison between the near-infrared and H$\alpha$ classifications of TTs.

The contribution of the circumstellar matter to the SED slope is often used to recognize different categories 
of young stellar objects (YSOs) by following an observational classification based on the near-infrared spectral 
index \citep{Lada84,Wilking89,Andre93}. 
This classification suggests a scenario for the evolution of YSOs, from Class 0 to Class III, 
which is well established for TTs. 

Although the present work is focused on HAeBes, 
it is useful to describe the meaning of the different TTs classes in order to illustrate how the spectral classification 
relates to disc structure and evolutionary sequence. 
We summarize here the phases of pre-main sequence evolution, based upon the review by \cite{Feigelson99} 
and following the schematic illustration by \cite{Carkner98}. 
The earlier evolutive phases are represented by Class 0 (infalling protostar) and Class I (evolved protostar).
Both contain objects that are surrounded by circumstellar shell and show increasing SED in the infrared range. 
Most of the Class II objects are CTTs, which are expected to have an accretion disc and to show flattened spectral index. 
The later phase is represented by Class III, in which are included WTTs and post-TTs. 
The disc is thin or non-existent in Class III objects, since they have little or no infrared-excess and show 
steeper slope of the SED. 

For HAeBes the spectral classification and its relation with an evolutionary sequence are less advanced than 
the TTs classes. 
In Sect. \ref{sectgroups} we present the HAeBes groups and a tentative scenario for their evolution,
as proposed by other authors.

\subsection{The HAeBes groups}
\label{sectgroups}

The flat disc geometry was also adopted to model the protoplanetary discs of HAeBes candidates detected 
by {\it PDS}. \cite{Sartori03} (hereafter SGH03) classified these stars in different groups, 
based on the SED shapes and the fraction of circumstellar luminosity,
expressed by $f_c = (L_{\rm tot} - L_{\star})/L_{\rm tot}$.
Adopting the spectral index $\beta_1 = 0.75 {\rm log}(F_{12}/F_V)-1$ 
\citep{Torres98}, the stars were classified as follows: the higher infrared excesses ($f_c>70\%$) are found 
in the $\beta_1 > 0$ group, which contains 44 per cent of the sample with increasing spectrum in the visible 
to infrared range; the $-1 < \beta_1 < 0$ group has 50 per cent of the stars showing intermediate amounts of 
circumstellar emission ($f_c =$ 10 to 70 \%); in the third group ($\beta_1 < -1$) there is 6 per 
cent of the stars having the lower values of infrared excess ($f_c < 10 \%$). The groups were 
established following the idea of an evolutionary scenario for HAeBes proposed by \cite{Malfait98}, 
similar to the TTs classes. In this scenario, values of $\beta_1 > 0.7$ correspond to embedded 
objects, like R CrA; and $\beta_1 < -1$ represents the stars showing more evolved discs, 
like Vega. The intermediary group has objects with SED similar to several known HAeBes. 

This classification is almost the same as the HAeBe groups presented by \cite{Hillenbrand92}, 
based on the slope of the infrared continuum. It is also comparable to the groups proposed by 
\cite{Meeus01} that studied a sample of 14 isolated Herbig stars. Their two major groups suggest 
different circumstellar geometries: the Group I sources are associated with an optically thin, flared 
region surrounding an optically thick disc, while Group II sources have a flat disc \citep{Meeus02}. 
One of the differences of this comparison is related to the SGH03 group of embedded objects. 
Since they cannot be considered isolated Herbig stars they were not classified by Meeus et al. 
Another difference is the subdivision based upon the presence (a) or the absence (b) of the 10 
$\mu$m silicate feature, which is not considered by SGH03. 
Table \ref{tabstars} summarizes (in the last columns) the comparison of the groups above described.

SGH03 estimated the circumstellar luminosity of the {\it PDS} HAeBes adopting the GH02 flat disc model. 
They verified that the fitting-quality was $gof > 0.4$ (a bad data fitting) for more than 30 per cent of the objects. 
This suggests that different disc geometry should be more appropriate for an important part of the sample.

\subsection{Main objectives}

We decided to adopt a flared configuration, by following the \cite{DDN2001} (hereafter DDN01) model 
for a passively irradiated circumstellar disc with an inner hole. Another improvement on our previous 
work is the method to optimise the parameters estimation based on genetic algorithms (GA). 
This automatic $\chi^{2}$ minimisation procedure was developed to increase the efficiency in the 
process of the SED fitting. 

The main goals of the present work are: 
(i) to describe the implementation of the GA method for the flared disc model; 
(ii) to discuss the derived disc parameters in a variety of SED shapes.
Once the GA method has been confirmed to be a valuable tool for SED fittings, the scope of the second
goal of this paper is to discuss the value of different spectral classifications in determining disc 
structures; the relevance of flat {\it versus} flared geometries; and on degeneracy of the parameters
derived from SED modelling.

We organized the structure of the paper in the following sequence. In Sect. \ref{sectmodel} we present
the adaptation of the DDN01 model, for which a new code was developed. A test of the rewritten routines was 
made by applying them to AB Aur. Sect. \ref{sectga} describes the GA method of parameters estimation and 
its implementation. Time efficiency and error estimative provided by the GA method are discussed in a comparison 
with our previous calculation procedure. In order to illustrate the application of the GA method to fit different 
SED shapes, using the DDN01 model, we present in Sect. \ref{secttest} the results obtained for some of the 
{\it PDS} stars. The sample was selected according different $\beta_1$ ranges to represent different 
groups of HAeBe stars. We choose to study only a few stars in this first presentation of the GA method 
just to verify if the adopted disc geometry is suited to fit different SED shapes. In a forthcoming paper 
we intend to apply the GA method to evaluate the circumstellar structure of the {\it PDS} stars that 
probably have flared discs. Considering the different levels of infrared excess shown by the selected 
sample, we discuss in Sect. \ref{sectdiscus} the circumstellar luminosity compared with the decreasing 
slope of the SED, which has been interpreted as a sequence for the disc evolution. 
The quality of the SED fitting is used to compare different morphologies assumed in the flared and the 
flat models.
For this purpose we describe in Appendix \ref{ttapp} the GH02 flat disc model.
Some inconsistencies of the derived parameters with respect to the observed ones are found for HD 141569,
leading us to discuss the degeneracy of the information obtained from SED modelling.
Finally, in Sect. \ref{sectconclus} the main conclusions are summarized. 


\section{The flared disc model}
\label{sectmodel}

\cite{Chiang97} developed a flared disc model that assumes hydrostatic radiative 
equilibrium for passive dust discs of TTs. DDN01 proposed a modified 
version of this model by including a puffed-up inner rim in the central disc region. This rim structure 
projects a shadow onto the disc and assumes that the disc's gaseous component is optically thin to 
stellar radiation. The dusty component is optically thick, when the accretion rate through the disc is very low. 

\begin{figure}
\includegraphics[width=2.3in,height=3in,angle=-90]{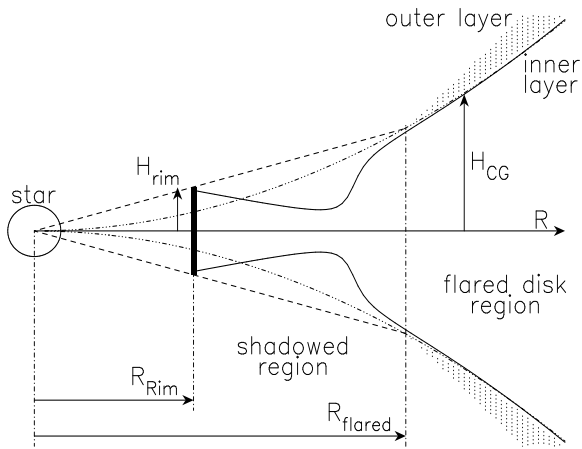}
\caption{Geometric scheme for the model adapted from DNN01.}
\label{figgeometric}
\end{figure}

To fit the SED we adopted the DDN01 model equations. For effects of calculation, the disc is divided 
into three parts: the inner rim, the shadowed region and the flared region (this has two layers: an 
illuminated hot-layer and an inner cold-layer). 
The first step is to establish a vertical boundary irradiated directly by the star, considering the effect 
due to shadowing from the rim, and the variations of scale height as a function of the radius. 
Figure \ref{figgeometric} presents a reference for the geometry of the model. 
The FitCGPlus disc parameters are: radius, $R_D$, mass, $M_D$, inclination, $\theta$, 
density power law index, $p$, and inner rim temperature, $T_{rim}$. The stellar parameters are: 
distance, $d$, mass, $M_{\star}$, luminosity, $L_{\star}$, and temperature, $T_{\star}$.

We are aware that the treatment of the shadow by the inner rim in the DDN01 model is rather simple, and that full 
3-D axisymmetric radiative transfer models of \cite{DD04} show that the shadow is in reality significantly less extreme.
However, for the SED this does not appear to make too much of a difference. Another drawback of the DDN01 
model compared to full radiative transfer models is the assumption that the inner wall radiation is a pure 
blackbody at the dust evaporation temperature: in 3-D radiative transfer calculations the inner wall is not 
a pure blackbody emitter and as a result produces slightly weaker near-infrared emission (Dullemond, 
private communication). These differences are not expected to strongly affect our results. 
The advantage of using the DDN01 model over full radiative transfer models is the speed of calculation of 
large numbers of models in a short time.

In the present work we keep the original nomenclature and symbols described in Sect. 2 of DDN01 paper,
which gives details on the physics of the model. C. P. Dullemond developed an interactive tool for applying 
the DDN01 model of protoplanetary discs. A public domain IDL code for their model, named FitCGPlus 2.1, 
is provided together with opacity tables for different grain composition, as well as photometric and spectral 
data of the star AB Aur, for testing. It is important to note that the IDL version does not reproduce the
same level of details of the stellar spectrum features (in the blue spectral region) as presented in 
Figure 8 of DDN01. However, our results are not affected by the stellar spectrum, since our 
interest is mainly related to the SED fitting in the infrared region.

Considering that our GA libraries are written using C++ language, and aiming to increase the
speed of calculation, we needed to rewrite the code for the DDN01 model.
The code (called HABdisk) was built following a bottom-up strategy, and the result of each 
intermediary routine and function was compared to their correspondent in FitCGPlus. 

\begin{figure}
\includegraphics[width=2.3in,height=3.1in,angle=-90]{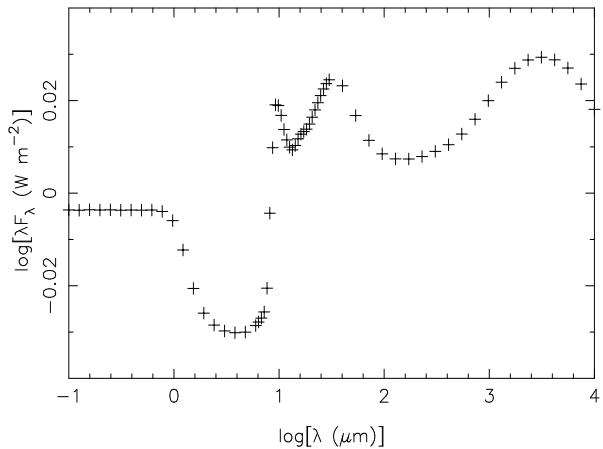}
\caption{Comparison of the FiTCGPlus with the HABdisk results for AB Aur.}
\label{figidlxc}
\end{figure}

The comparison of the FiTCGPlus synthetic flux with HABdisk result gives a mean ratio of 
[$( \lambda {\rm F}{_\lambda} )_{\rm FiTCGPlus} -
( \lambda {\rm F}{_\lambda} )_{\rm HABdisk} ] / ( \lambda {\rm F}{_\lambda} )_{\rm HABdisk} \sim 1.3\%$, 
that seems to be due to binary computations during iterations. 
Figure \ref{figidlxc} illustrates these small differences by showing the deviations for each considered wavelength.

\begin{figure*}
\centering
\hskip -3in \includegraphics[width=3.3in]{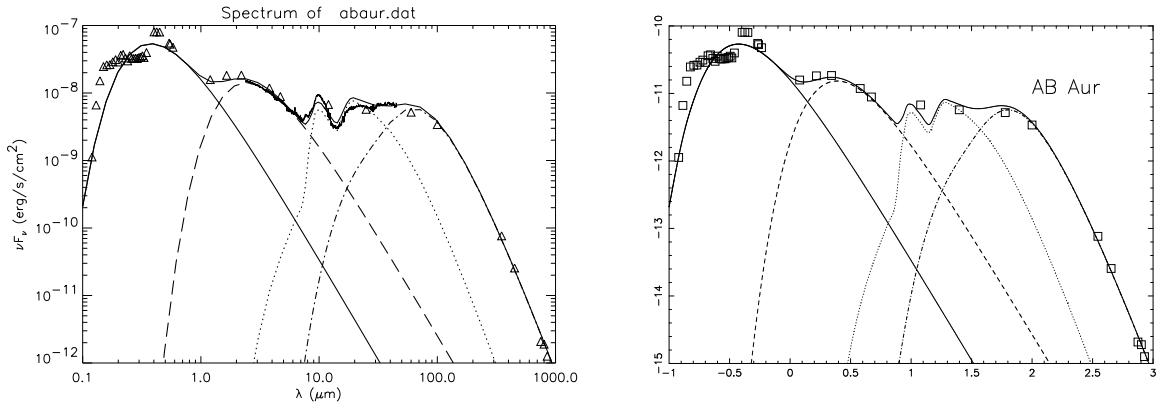}
\vskip -2.31in \hskip 3.3in \includegraphics[height=2.63in,angle=-90]{abaurcpp.eps} \vskip 0.2in
\caption{AB Aur synthetic SEDs obtained with two codes. 
The continuous thin line represents the star emission; dashed line represents the rim emission; 
dot-dashed line for the disc cold-layer; and dotted-line for the disc hot-layer.
(left) FitCGPlus, the IDL version of DDN01 model; 
(right) HABdisk, the C++ version of DDN01 model. 
The plot is given in ${\rm log}\left[ \lambda {\rm F}{_\lambda} \left({\rm Wm}^{-2} \right) \right]$ 
versus ${\rm log}\left[ \lambda \left( \mu {\rm m} \right) \right]$. }
\label{figabaur}
\end{figure*}

The HABdisk code was applied to AB Aur (a DDN01 canonical model) using the parameters from \cite{DDWW03}:
$M_{\star}=2.5$ M$_\odot$, $L_{\star} = 47$ L$_\odot$, $T_{\star}=9750$ K, $M_D=0.1$ M$_\odot$, 
$R_D = 400$ AU, $\theta = 65^{\rm o}$, $T_{rim} = 1500$ K, and $D$ = 144 pc.
Figure \ref{figabaur} presents a comparison of the SEDs calculated with the codes FitCGPlus and HABdisk.

\section{The GA method}
\label{sectga}

This section explains how the GA method was implemented to the DDN01 model 
(more details can be found in the Appendix \ref{gaapp}).

Considering that the GA nomenclature can be misplaced in the astrophysical context, we first present the 
translation from one field to another. A {\it parameter} (e.g. disc radius), corresponds to the concept of a 
``gene'', and a {\it change in a parameter} is a ``mutation''. A {\it parameters set} that yields to a possible 
solution corresponds to a ``chromosome''. 
An ``individual'' is a {\it solution}, that is composed by one parameters set and two additional GA control variables. 
One of these variables is $\chi^2$, which means the 
``adaptation'' level. The other control variable is $\Phi$, the genetic operator described in Sect. \ref{sectopers}. 
The term ``generation'' means ``all the individuals'' (or {\it all the solutions}) present in a given iteration. 

\subsection{Implementation}

The HABdisk code uses the same parameters of FitCGPlus, which are: $M_{\star}$, $R_D$, 
$\theta$, $M_D$ and $p$. On the other hand, we decided to allow $T_{rim}$ to be variable. 
The stellar parameters $d$, $L_{\star}$ and $T_{\star}$, 
are adopted from observations (see Sect. \ref{sectinputpar}).

Essentially, the GA method presented herein implements a $\chi^2$ minimisation of the SED fitting 
provided by the DDN01 model. There are three main advantages in using GA for this task: 
(i) the GA method potentially browses the whole permitted parameter space, better avoiding the 
``traps'' of local minima; (ii) the method is not affected by changes in the model; 
(iii) the GA implementation does not need to compute the derivatives of $\chi^2$ 
(like $\partial \chi^2 / \partial R_D$ for example) required by the usual methods. 
This fact simplifies the code and minimises computer errors due to gradient calculations.

\subsubsection{Data structures and control variables}

The main structures used to manipulate the data are linked lists containing the 
solutions (parameters set, the adaptation level, $\chi^2_i$, and the genetic operator, $\Phi_i$), expressed by 
\begin{equation} 
S_i =\left[ \left(R_D{}_i, \theta_i, M_D{}_i, p_i, T_i \right), \left(\chi^2_i, \Phi_i \right) \right]
\end{equation}
where $S_i$ means the $i$th solution, and $T_i$ is the $i$th $T_{rim}$.

Following \cite{Goldberg89} the code starts with the construction of the first generation, where all 
parameters are randomly chosen within an allowed range (for example, $50 \leq R_D \leq 1000$ AU). 
In the present work, the number of different parameter sets in the first generation is assumed to be 100.
In the next step the evaluation function runs the DDN01 model for each solution, and compares the 
synthetic SED with the observed data to find the $\chi^2$, using the expression given by \cite{Press95}: 
\begin{equation}
\chi^2_i =\frac{1}{N}\sum\limits_j^N {\left( {F_j - \phi_{ij} } \right)^2} 
\end{equation}
where $N$ is the number of observed data points, $F_j$ is the observed flux at wavelength 
$\lambda_j$ and $\phi_{ij}$ is the calculated flux for the solution $S_i$. 
The smallest $\chi^2$ corresponds to the goodness-of-fitting, or $gof$.

\subsubsection{The genetic operators}
\label{sectopers}

As the evaluation function has been applied to all solutions, the judgement procedure sorts the list by 
increasing $\chi^2$, and sets to $\Phi_i$ one of the genetic operators: copy, crossover
(also called recombination), mutation or termination. Each $\Phi$ is attributed to a fraction 
of the generation following the values suggested by \cite{Koza94}, \cite{Bentley02} and references therein. 

The next generation is constructed by applying the correspondent genetic operators. The copy uses 
an elitist selection, since the solutions with the smallest $\chi^2_i$ are copied to the next generation. The 
crossover operator randomly mixes parameters from two different solutions. For example:
$S_a = \left[R_D{}_a, \theta_a, M_D{}_a, p_a, T_a \right]$ and 
$S_b = \left[R_D{}_b, \theta_b, M_D{}_b, p_b, T_b \right]$ can give
$S_c = \left[R_D{}_b, \theta_a, M_D{}_a, p_b, T_a \right]$. 
Mutation reproduces the same original solution, except for one of the parameters that is randomly changed.

\begin{figure}
\includegraphics[width=2.1in,height=3.3in,angle=-90]{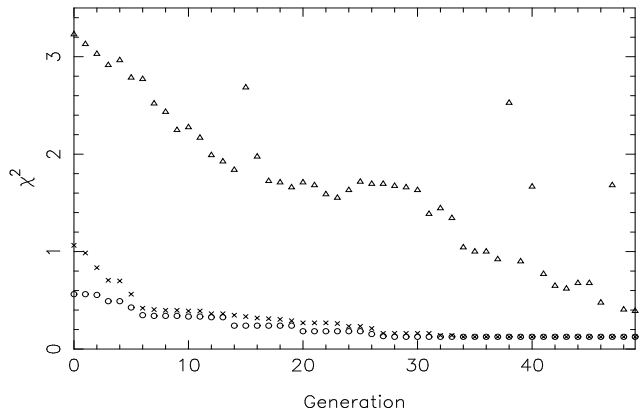} 
\caption{The distribution of $\chi^2$ for a sequence of generations. The circles are used to represent 
the best solution in each generation, and the triangles represent the worst ones. 
the 5\% best solutions are found below the level indicated by the crosses.}
\label{figgenerations}
\end{figure}

The method continues to build new generations until the end condition is reached, as illustrated 
by the schematic view shown in Figure \ref{stepsGA}. The distribution of $\chi^2$ for a sequence 
of 50 generations in the SED fitting of AB Aur is presented in Figure \ref{figgenerations}.

\subsection{Tests and error estimative}

\subsubsection{Comparing to previous method}
\label{sectcomparing}

\begin{figure}
\centering{\includegraphics[width=2.0in,height=3in,angle=-90]{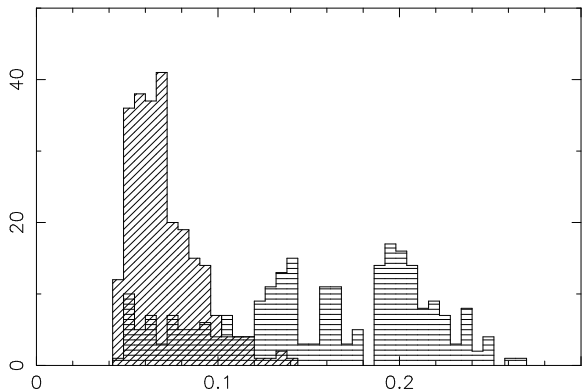}}
\caption{Histograms of $\chi^2$ values, comparing two different minimisation methods used to fit the 
SED of AB Aur. The Newton-Raphson method results are filled with horizontal hatching, while the GA 
method results are filled with $45^{\rm o}$ hatching. }
\label{fighistogof}
\end{figure}

Figure \ref{fighistogof} shows a comparison between the GA method and the classical Newton-Raphson 
method adopted in our previous work (GH02), both used in the SED fitting of AB Aur. 
The test is limited to 256 runs, each one starting from random values for the free parameters. 
The number of iterations is fixed, aiming to obtain a time independent comparison. 
The GA method provides a single concentration that approaches the global minimum around 
$\chi^2 = 0.046$. The spurious peaks produced by the classical method are due to the several 
local minima that usually result from methods based on gradients. We also evaluated the time 
efficiency and verified that the GA method is about 16\% faster.

\subsubsection{Confidence levels}

An analysis of the $\chi^2$ behaviour as a function of the parameters variation is used to estimate error bars, 
determining the confidence levels given by \cite{Press95}, as described in Appendix \ref{errorapp}. 
The error bars estimated for 1$\sigma$ confidence level and respective disc parameters for AB Aur are 
$M_D=0.1 \pm 0.004$ ${\rm M}_\odot$, $R_D = 400 \pm 44$ AU, $\theta = 65 \pm 3^{\rm o}$, and 
$T_{rim} = 1500 \pm 26$ K. These results are in agreement with another error bars estimation provided 
by the surface contour levels.

\begin{figure}
\centerline{\includegraphics[width=2.13in,height=3in,angle=-90]{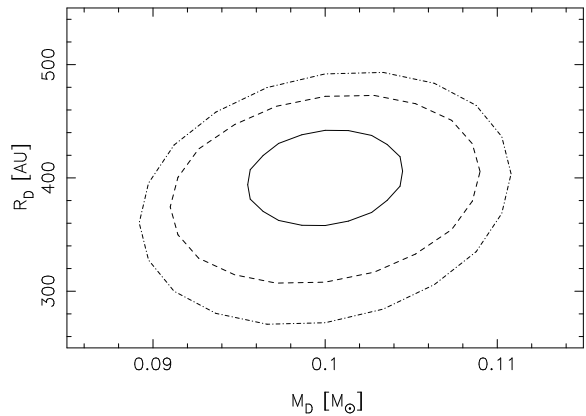}}
\caption{$gof$ contour levels in parameter space projected in [$M_D$, $R_D$] plane evaluated for 
AB Aur. Levels are $\chi^2=0.082$ (continuous line), $\chi^2=0.15$ (dashed) and $\chi^2=0.21$ (dot-dashed).}
\label{figlevels}
\end{figure}

Figure \ref{figlevels} presents the contour levels of the $gof(M_D, R_D)$ surface, which was calculated 
for a set of 400 random pairs of disc mass and radius around the parameters for AB Aur model taken 
from \cite{DDWW03}, resulting a mean minimum $gof \sim 0.046$. 
This means that the GA method converged to a narrow range around the same parameter set obtained by them.
The three contours represent the confidence levels for 68\%, 90\% and 99\%.


\section{Tests with {\it PDS} stars}
\label{secttest}

\subsection{Sample Selection and observational data}

Motivated by the efficiency of the GA method, in a further work we intend to model the protoplanetary 
discs of the 108 HAeBe detected by {\it PDS}, which are listed by \cite{Vieira03}. The worth in evaluating 
the circumstellar luminosities of a large sample of objects is to compare them with spectral and polarimetric 
characteristics, providing a statistical analysis to be presented by our group in a forthcoming paper. 
The main goal of the present work is to apply the GA method for a few objects only, in order to verify 
the quality of the fitting for objects showing different SED shapes and different levels of infrared excess. 
We decided to choose four stars to represent the groups described in Sect. \ref{sectgroups}, selected by the slope
of their SED in the near-infrared. According \cite{Ancker01}, the infrared excess in HBe 
stars is due to a spherical dusty envelope, while HAe stars are better represented by a thick edge flared disc. 
This suggestion motivated us to select A type or late-B type stars from the {\it PDS} sample to illustrate 
the application of the SED fitting procedure described in Sect. \ref{sectga}. Table \ref{tabstars} presents the main 
characteristics of the stars considered representatives of different ranges of $\beta_1$ discussed by SGH03. 

\begin{table*}
\centering
\caption{List of {\it PDS} stars selected as representatives of different $\beta_1$ groups (SGH03). 
Spectral types, estimated visual extinctions and stellar parameters are also given. A comparison with HAeBe 
groups suggested by \protect\cite{Meeus01} and \protect\cite{Hillenbrand92} are presented respectively in the two last columns. 
IRAS 07394-1953 (PDS257) is not classified in the groups of Meeus et al., since it cannot be considered 
an isolated Herbig star.}
\begin{tabular}{@{}llccccccccc@{}}
\hline
\raisebox{-1.50ex}[0cm][0cm]{PDS} & \raisebox{-1.50ex}[0cm][0cm]{Name} &
\raisebox{-1.50ex}[0cm][0cm]{S.T.} & $A_V$ & $T_{\star}$ & $L_{\star}$ & $d$ &
\raisebox{-1.50ex}[0cm][0cm]{$\beta _1$} & \raisebox{-1.50ex}[0cm][0cm]{SGH03} & Meeus & Hillenbrand \\
& & & (mag) & (K)& (L$_\odot$) & (pc) & & & group& group \\
\hline
398 & HD 141569 & B9.5Ae & 0.25 & 10000 & 20 & 106 & -1.74 & $\beta _1<-1$ & II & -- \\
022 & BD-14 1319 & A0 & 0.37 & 9760 & 78 & 800 & -0.42 & $-1<\beta_1< 0$ & I & I \\
130 & IRAS 06475-0735 & A & 1.9 & 10140 & 35 & 995 & 0.23 & $0<\beta_1< 0.7$ & I & II \\
257 & IRAS 07394-1953 & A0 & 2.7 & 10000 & 150 & 4000 & 0.74 & $\beta_1>0.7$ & -- & II \\
\hline
\end{tabular}
\label{tabstars}
\end{table*}

There is a lack of information about IRAS 07394-1953 and IRAS 06475-0735, being only previously 
studied by the {\it PDS} group \citep{PDS1, Vieira03}. BD-14 1319 has been included in a polarimetric 
study by \cite{Yudin98} but no polarization was detected for this star. Complementary photometry and 
low-resolution spectroscopy are reported by \cite{Mirosh99}. 

On the other hand, HD 141569 is a very well studied object, showing a circumstellar structure that has 
been considered similar to the disc of $\beta$ Pictoris. 
The B9.5Ve primary and two T Tauri stars form a triple system, separated by 9 arcsec, with age of 5 Myr, 
as estimated by \cite{Weinberger00}. \cite{Merin04} estimate $L=22$ L$_\odot$ for the primary. 
Based on the study of circumstellar accretion of a sample of {\it PDS} HAeBe stars, \cite{Guimaraes06} 
have determined stellar parameters, metallicity and rotational velocity. 
No evidence of accretion episodes was found for HD 141569, probably related to an evolved status of the disc.

 A NICMOS image of the circumstellar structure was obtained by \cite{Weinberger99} showing a disc size 
of almost 5 arcsec (500 AU in projection). Observations of the disc in scattered light at near-infrared and
optical ranges revealed a complex dust distribution showing a large inner hole out to 150 AU from the 
central star and two rings at 185 AU and 325 AU \citep{Augereau99,Weinberger99,Mouillet01,Boccaletti03,Clampin03}. 

The remnant mass of H$_2$ was estimated as 20--460 M$_{Earth}$ by \cite{Zuckerman95} based on CO 
measurements around HD 141569. More recently \cite{Jonkheid06} have studied the chemistry and gas 
temperatures in the disc using a radiative transfer code to fit the observed CO emission lines. 
In order to reproduce the data, they adopted a power-law gas distribution of 80 M$_{Earth}$ starting at 80 AU
from the central star.

The optical photometry was obtained from {\it PDS} and far-infrared from {\it IRAS} Point Source 
Catalogue. An upper limit millimetric photometry at 1.1 mm, obtained by \cite{Sylvester96} is available only for 
HD 141569, which gives a constraint for the cold layer disc component. Near-infrared data for 
IRAS 06475-0735, IRAS 07394-1953, and BD-14 1319 are from {\it 2MASS} catalogue, 
and HD 141569 from \cite{Sylvester96}. 

\subsection{Input parameters}
\label{sectinputpar}

The optical and near-infrared data were corrected for interstellar extinction by using the normal relation 
$A_V = 3.1 E(B-V)$ given by \cite{Savage79} and extinction law $A_{\lambda}$/$A_V$ from \cite{Cardelli89}. 
Considering the UV excess presented by Herbig stars, the ($B-V$) excess was estimated by adopting 
$E(B-V) = E(V-I)/1.6$ from \cite{Schultz06}. The intrinsic colours and bolometric corrections from \cite{Bess98} 
were adopted in the estimation of $A_V$ and absolute bolometric magnitude. \cite{Vieira03} have listed the 
association of the objects with the star forming regions CMaR1 (1000 pc, IRAS 06475-0735), 
NGC2149 (800 pc, BD-14 1319), and G236.4+1.49 (4300 pc, IRAS 07394-1953). 
For these stars we have adopted the distances of the respective clouds and assumed an error of $10\%$. 
Hipparcos parallax has been obtained for HD 141569 giving a better determination for distance 
($99 \pm 10$ pc) in this case. 

Temperatures were estimated from the $T_{eff}$ - Spectral Type relations given by \cite{deJager87}, 
assuming a deviation of $T_{\star} \pm 200$ K. 

Table \ref{tabstars} gives the distance, luminosity and temperature of the stars that are fixed input parameters. 
The free parameters and respective adopted limits are: 
stellar mass (2 to 20 ${\rm M}_\odot$);
disc mass (0.001 to 0.5 M$_\odot$); 
disc radius (100 to 1000 AU); 
and inclination angle (0$^{\rm o}$ to 70$^{\rm o}$). 
The only exception is HD 141569, for which we have fixed the disc size as 500 AU. 

Two of the model parameters are not fitted by us in the same way as made by DDN01 and \cite{DDWW03}.
The slope of the power-law $\Sigma (R) = \Sigma_0 R^p$, which describes the surface density, is assumed 
not to be a completely free parameter, since only the discrete values $p$= -2, -1.5 or -1 were used. On the 
other hand, the disc inner temperature is not fixed at 1500 K. Instead of fitting the inner disc radius, we adopted 
$T_{rim}$ as a free parameter, ranging from 300 to 2000 K, in order to have different sizes of the inner disc hole. 

The dust opacity given by \cite{Draine84} is adopted, assuming a 100\% silicate grain composition.
It must be kept in mind that different grain compositions cause significant 
degenerations in the model solutions (see Sect. \ref{sectquality}). 
We are not attempting to fit in detail the spectral features from the near- to the far-infrared,
considering the lack of detailed observations, like {\it ISO} data, for most of the {\it PDS} stars. 
For this reason, we decided to adopt a standard grain composition for the whole sample.

\subsection{Results}

\begin{figure}
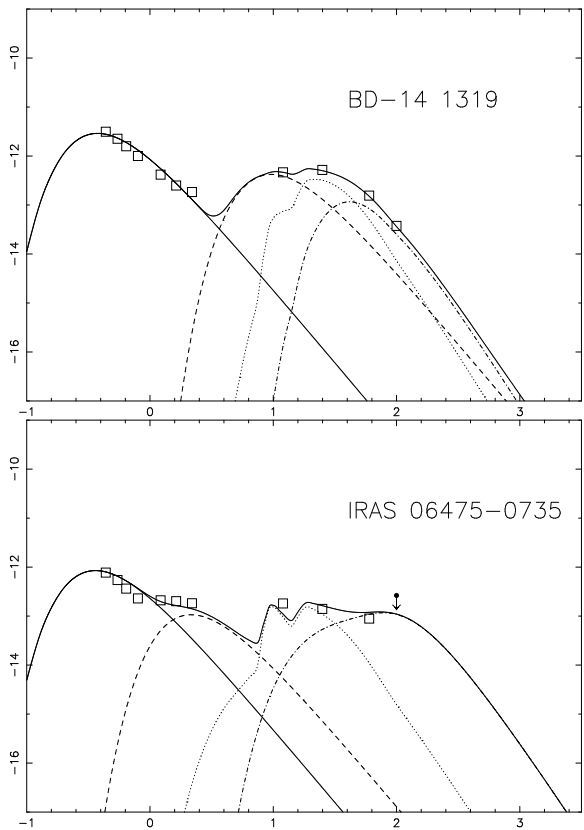

\centerline{\includegraphics[width=2.14in,height=3in,angle=-90]{pds022.eps}}
\centerline{\includegraphics[width=2.14in,height=3in,angle=-90]{pds130.eps}}
\caption{Synthetic SED obtained for the two objects having intermediary values of $\beta_1$
The plots are given in ${\rm log}\left[ \lambda {\rm F}{_\lambda} \left({\rm Wm}^{-2} \right) \right]$ 
versus ${\rm log}\left[ \lambda\left(\mu {\rm m} \right) \right]$. } 
\label{pds022-130}
\end{figure}

\begin{figure}
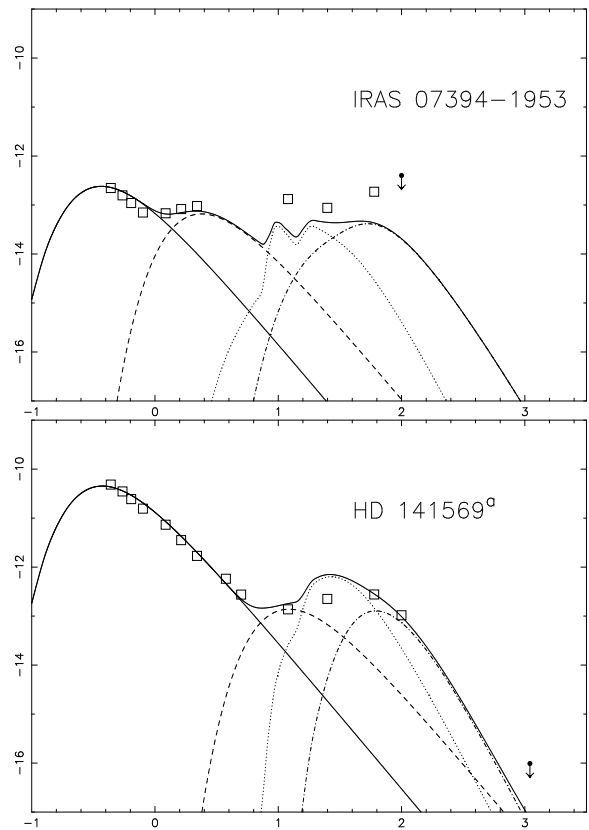

\centerline{\includegraphics[width=2.14in,height=3in,angle=-90]{pds257.eps}}
\centerline{\includegraphics[width=2.14in,height=3in,angle=-90]{pds398a.eps}}
\caption{Synthetic SED obtained for object showing extreme values of $\beta_1$: 
IRAS 07394-1953 ($\beta_1 = 0.74$) and HD 141569 ($\beta_1 = -1.74$). 
The bad fitting quality in the infrared region is discussed in Sect. \ref{sectquality}. } 
\label{PDS257-398}
\end{figure}

Table \ref{tabparams} presents the parameters derived from SED fitting by the HABdisk code, and 
the fraction of circumstellar luminosity ($f_c$).
A very good quality of fitting was achieved for BD-14 1319 and IRAS 06475-0735, as shown in Figure \ref{pds022-130}. 
The same cannot be said for the other two objects displayed in Figure \ref{PDS257-398}. 

\begin{table*}
\centering
\caption{Derived model parameters for the sample, where $f_c$ gives the per cent amount of circumstellar luminosity.
The best fitting of HD 141569 is provided by parameters set (b). }
\begin{tabular}{@{}llcccccccccc@{}}
\hline
PDS & Name & &$M_{\star}$ (M$_\odot$) & $R_D$ (AU) & $M_D$ (M$_\odot$) 
& $T_{rim}$ (K) & $\theta$ ($^{\rm o}$) & $p$ & $gof$ & $f_c$ ({\%}) & $\beta_1$ \\
\hline
\raisebox{-1.50ex}[0cm][0cm]{398} & \raisebox{-1.50ex}[0cm][0cm]{HD 141569} 
 & (a) & 2.4 & 500 & 0.002 & 303 & 0.6 & -2.0 & 0.021 & 1.4 & -1.74 \\
 & & (b) & 2.4 & 13 & 0.06 & 1085 & 0.6 & -2.0 & 0.006 & 1.4 & -1.74 \\
022 & BD-14 1319 & & 2.8 & 690 & 0.003 & 380 & 40 & -1.0 & 0.006 & 21 & -0.42 \\
130 & IRAS 06475-0735 & & 2.0 & 309 & 0.20 & 1705 & 53 & -1.5 & 0.016 & 32 & 0.23 \\
257 & IRAS 07394-1953 & & 2.0 & 859 & 0.64 & 1838 & 47 & -2.0 & 0.098 & $>35$ &0.74 \\
\hline
\end{tabular}
\label{tabparams}
\end{table*}

\begin{figure}
\centerline{\includegraphics[width=2.2in,height=3in,angle=-90]{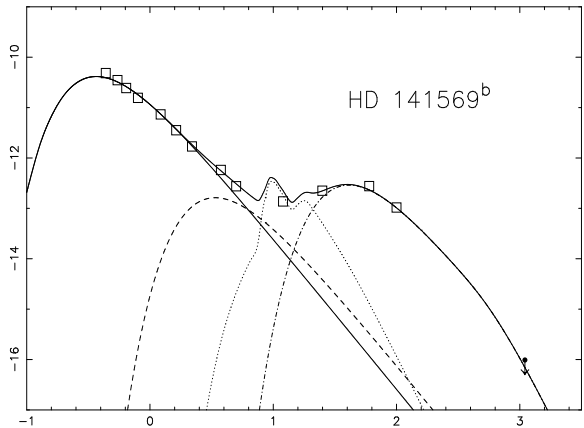}}
\caption{Best SED fitting obtained for HD 141569, which gives $R_D=13$ AU (see text). }
\label{figcurto}
\end{figure}

Considering the low quality of the fitting obtained for HD 141569 when fixing $R_D = 500$ AU, we decided
 to test a second HABdisk run for this star, and to compare different sets of parameters. In this case, 
an excellent SED fitting was obtained (see Figure \ref{figcurto}), but the derived disc parameters are 
unrealistic, as shown in Table \ref{tabparams} (see also Sect. \ref{secthd141569}).

\cite{DDWW03} have discussed the variety of the spectra of isolated HAeBes, members of groups I and II of 
\cite{Meeus01}, in the context of the DDN01 model. They found good SED fittings for both groups, especially 
in the case of IIa objects. However, some of the fitting resulted in compact disc radii that are inconsistent with 
the typical sizes measured around young stars. They concluded that while the DDN01 model is capable to 
fit the SED of Group IIa objects, the derived model parameters seem difficult to understand. 


\section{Discussion}
\label{sectdiscus}

\subsection{The quality of the fittings}
\label{sectquality}

In this section we first discuss the results of goodness-of-fitting and the adequateness of the DDN01 
model for the studied objects. It is confirmed that the flared geometry is well suited for the stars showing 
intermediary values of $\beta_1$ index, while the same cannot be said for the objects with extreme 
SED slope ($\beta_1 < -1$ or $\beta_1 > 0.7$).

Despite the good fittings, only IRAS 06475-0735, which shows a flattened SED similar to AB Aur, 
had more realistic derived parameters. For BD-14 1319, an unrealistic $T_{rim}$ was required 
to fit its double peaked SED. A large central gap yielding to a cooler inner disc temperature, as found by
 \cite{Alencar03} in the SED fitting of AK Sco for example, could be the reason. However, the derived 
$T_{rim}$ is much lower than the usually adopted dust evaporation temperature. The absence of a puffed-up rim 
seems to be a better explanation, since this star does not present a bump in the near-infrared. 

It can be noted from Figure \ref{PDS257-398} that the object showing the largest infrared emission, 
IRAS 07394-1953, could not be fitted by the adopted model. 
A possible reason is contamination by nebulosity or cirrus as indicated by the high excess at 60 
$\mu$m and 100 $\mu$m verified on the IRAS-ISIS images, which respectively show flux densities 
$F_{60} >$ 5 MJy/sr and $F_{100} >$ 20 MJy/sr nearby the location of IRAS 07394-1953. 
For the other fields, lower levels are found ($F_{60} \leq 3$ MJy/sr and $F_{100} \leq 10$ MJy/sr), 
leading us to suggest that IRAS 07394-1953 is the most embedded object of the sample.
Probably in this case, a circumstellar shell component is needed to reproduce the far-infrared excess, 
as discussed in the comparison of different models (see below). 
Only a lower limit of $f_c >35\%$ could be estimated for this star since the infrared data were not fitted. 
On the other extreme, HD 141569 does not have a good fitting if the observed disc size is adopted 
(see Sect. \ref{secthd141569}). 

In order to verify the question of whether a flat disc model would provide good SED fittings for the HAeBes 
studied in the present work, we also present previous results obtained with the GH02 model, 
which is described in Appendix \ref{ttapp}. 
Table \ref{tabapptt} gives the parameters derived for this sample, showing that the quality of the fitting 
($gof$) is worst than that obtained with HABdisk, except for IRAS 07394-1953. 
In this case, the spherical component included in the flat disc model could provide a better fitting 
of the far-infrared data. 
Independently of the adopted circumstellar geometries, the poor {\it gof}s achieved with the flat disc 
model seems to be related to the minimisation procedure, as discussed in Sect. \ref{sectcomparing} 
and shown in Figure \ref{fighistogof}.

Since a flat disc would be expected to explain the HD 141569 circumstellar structure
($\beta_1 < -1$, Group II of Meeus et al. 2002) we compare here the flat disc parameters 
with the HABdisk results, obtained for this star.
The disc size and the temperature resulted in reasonable values using the flat disc model,
but the system is seen almost edge-on, contradicting the observations.
Probably, this is due to the high disc luminosity produced 
by the optically thick disc assumed in the GH02 model, as expressed in Eq. \ref{equSd}.

The differences and similarities resulting from both models demonstrate that degenerated information
can also be derived from SED modelling.
In this case, there is no argument to distinguish between HABdisk and the GH02 flat disc model, 
when applied to HD 141569. 

\begin{figure}
\centerline{\includegraphics[width=2.2in,height=3in,angle=-90]{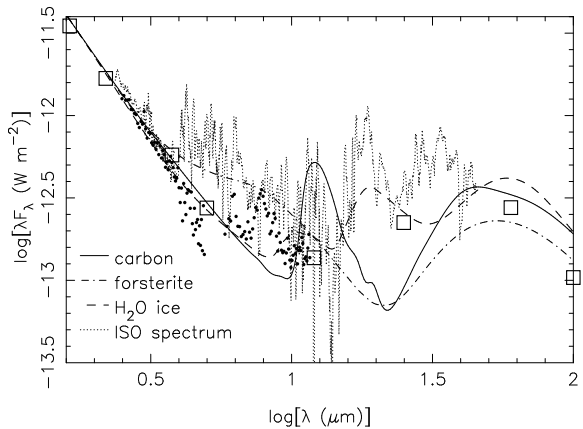}}
\caption{
HD 141569 {\it ISO} photometry (filled circles) from SS detector (2.5--4.9 $\mu$m) and 
SL detector (5.8--11.6 $\mu$m), and SWS spectroscopy. 
Several curves are used to represent the SED fittings obtained with different chemical compositions.}
\label{iso398}
\end{figure}

Another consideration regarding the problem of degeneracy is to verify how much grain composition 
affects the synthetic spectrum. 
For this study we have inspected the features shown by {\it ISO} 
photometry and spectroscopy available for HD 141569 (public archival data). 
The details in the spectral range 2 to 50 $\mu$m are presented in Figure \ref{iso398} that compares the {\it ISO} data 
with the SED calculated with the GA method by using different grain compositions. 
Table \ref{tabchimie} presents the obtained disc parameters and $gof$s. 
Among the three grain types, carbon and forsterite gave the worst $gof$s, when compared to the results 
for HD 141569 in Table \ref{tabparams}.
On the other hand, H$_2$O ice provided a good quality 
of fitting, but the mid- and far-infrared data were not well reproduced. 
The derived parameters are in the range defined by the results (a) and (b) presented in 
Table \ref{tabparams} indicating that the conclusions about HD 141569 are not biased by the choice of using 
astronomical silicates in the modelling.

\begin{table}
\centering
\caption{Disc parameters obtained with the SED fitting of HD 141569 by adopting different grain compositons.}
\begin{tabular}{@{}lcccc@{}}
\hline
 & $R_D$ (AU) & $M_D$ (M$_\odot$) & $T_{rim}$ (K) & $gof$ \\
\hline
carbon & 21 & 0.012 & 1050 & 0.1 \\
forsterite & 24 & 0.012 & 748 & 0.04 \\
H$_2$O ice & 65 & 0.010 & 576 & 0.01 \\
\hline
\end{tabular}
\label{tabchimie}
\end{table}

\subsection{The correlation of $f_c$ and SED slope}

The four objects of the sample show a correlation of circumstellar luminosity with spectral index $\beta_1$,
as can be noted in the two last columns of Table \ref{tabparams}. 
This correlation defines a distribution of the objects, where 
IRAS 07394-1953 ($\beta_1= +0.74$, $f_c > 35\%$) is in the beginning
and HD 141569 ($\beta_1 = -1.74$, $f_c = 1.4\%$) is in the end of the sequence. 
The diminishing of $f_c$ with the SED slope is in agreement with the HAeBe evolutionary sequence suggested 
by \cite{Malfait98}.

In a comparison with the circumstellar geometries of the two major HAeBes groups proposed 
by \cite{Meeus02}, the low $f_c$ of HD 141549 is consistent with their suggestion of a flat disc structure 
for Group II, while the intermediary $f_c$ values for BD-14 1319 and IRAS 06475-0735 are consistent 
with the optically thick disc surrounded by a flared thin region, as proposed for their Group I. 
On the extreme is IRAS 07394-1953, showing the highest $f_c$ of the sample, 
which precedes the two groups of Meeus et al. in the sequence, since it is not an isolated HAeBe.
However, there are no clear differences among the derived model parameters for these stars that were chosen
 to represent different groups, or SED shapes. Based only on the SED fitting it is difficult to distinguish the 
objects in an evolutionary sequence, mainly for those showing $-1 < \beta_1 < 0.7$.

\begin{figure}
\centerline{\includegraphics[angle=-90,width=3.1in]{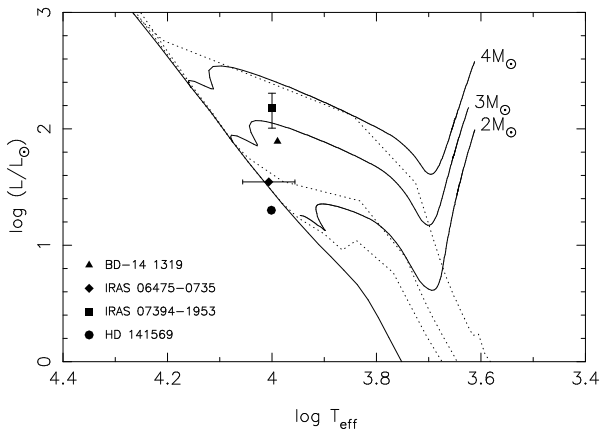}}
\caption{H-R diagram showing the position of the stars compared with the ZAMS (continuous line) and 
10, 5 and 1 Myr isochrones (dotted lines) from \protect\cite{Siess00}. Only the largest error bars are presented. }
\label{fighr}
\end{figure}

Indeed, the comparison of the stellar parameters on the H-R diagram shows no evidence 
of differences on their evolutionary status. 
In Figure \ref{fighr} the bolometric luminosities and effective temperatures of the four 
stars are compared with the isochrones provided by \cite{Siess00}. 
The Zero Age Main Sequence (ZAMS) and isochrones for 10, 5 and 1 Myr are displayed. 
Even considering the well-determined age of HD 141569 in the literature, it is not possible to infer 
the ages of the other objects in Figure \ref{fighr} due to the imprecision on 
distance or temperature determination, as indicated by the illustrative error bars. 
In spite of the age uncertainties, it can be noted that high values of $\beta_1$ 
correspond to higher positions above the ZAMS, but a large number of objects is needed to verify this trend.

In fact, the only hypothesis for the studied sample is the suggestion regarding the extremities of a possible sequence. 
$\beta_1 < -1.0$ objects showing very little circumstellar emission ($f_c < 10\%$), 
as the example of HD 141569, are the best candidates to have more evolved discs, 
while IRAS 07394-1953 probably is the opposite case. 
Any other inferences about the evolutionary status of objects 
require a demonstration for a large sample that we intend to develop in a further work. 

\subsection{HD 141569}
\label{secthd141569}

For HD 141569 the SED could not be well fitted with the disc radius measured from NICMOS image, 
indicating that the flared disc model fails to give the best fitting mainly in the mid-infrared region around 
$\sim 25$ $\mu$m. 
This is not due to a failure of the GA method to find the real best solution. 
It is verified that a different parameters set can provide an adequate SED fitting, but resulting in 
disc radius $R_D = 13$ AU.
While this seems to contradict the near-infrared observations, which revealed a disc size 
larger than 300 AU, it must to be noted that only sub-$\mu$m sized grains are seen in scattered light
at large distances.
The outer parts of the disc do not comprise the bulk of the disc mass, which is mostly concentrated closer to the star.

Probably, the evolved disc (or thin disc geometry) of HD 141569 is better explained by a flat structure instead. 
Our previous results obtained using the flat disc model 
are consistent with the observed size (see Appendix \ref{ttapp}).
However, that simple model also fails to reproduce the complicated structure of the inner disc of HD 141569
that seems not to be only related to differences on circumstellar geometry or degeneracy of the models. 
We trace below some information about grain sizes and chemical composition,
probably related to evolutionary aspects.

The presence of large grains \citep{Boccaletti03} and PAHs \citep{Sylvester96, Weinberger04, Geers06} 
have been reported for HD 141569, while silicate emission is not present \citep{Weinberger00}.
This can be verified in the spectral features presented in Figure \ref{iso398}. 
The {\it ISO} data show an almost flat SED near 10 $\mu$m, 
while a low intensity PAH emission is found at 8.6 $\mu$m. 

The absence of the 10 $\mu$m feature has been discussed by \cite{Meeus02} as a possible lack of 
small silicate grains, or a temperature effect that leads to a lack of small, hot silicates. According DDN01, 
the shadowing caused by the rim can suppress the strength of the silicate emission. \cite{Hernandez07} 
suggest that grain growth and/or settling, and transitions objects, which are stars with inner gaps in their 
discs \citep{Sicilia06}, are a possible combination to explain evolved discs.
The correlation between the mid-infrared and the sub millimetre slope, found by \cite{Acke04}, 
indicates that as the grains grow the disc structure evolves from flared to geometrically flat \citep{Dullemond02},
which seems to be the case of HD 141569. 

Another implication on the radial structure of this disc can be also related to dynamical processes.
Scattered light images of the optically thin dust disc were used by \cite{Augereau99} and \cite{Augereau04} 
in the investigation of the properties and the dynamics of the HD 141569 triple system. 
Their suggestions on different dust populations required to fit the full SED, as well as the processes 
of disc truncation, indicate a scenario too complex to be reproduced by the two models that we have studied. 


\section{Summary and Conclusions}
\label{sectconclus}

The goal of this paper is to describe a method to fit the SED of pre-main sequence stars, using the code 
called HABdisk, which was adapted from the flared disc model proposed by DDN01. 
The procedure uses the GA method in order to find the best parameters set, based on minimum $\chi^2$ criterion.
The tests applied to AB Aur confirm that the GA method provides a good estimation of the circumstellar parameters, 
reproducing a canonical DDN01 model. 

Aiming to provide an illustrative application of HABdisk, we selected four {\it PDS} stars 
that have similar mass and effective temperatures, but show different amounts of infrared excess,
as indicated by their spectral indices. 

\subsection{The value of the different spectral classification in determining disc structures}
\label{sectvalue}

The estimated circumstellar luminosities are in agreement with the expected correlation of SED shape with $f_c$,
which vanishes as the spectral index in the infrared decreases. 
The $f_c$ behaviour in the sample is consistent with the Groups I and II defined by \cite{Meeus02}, 
as summarized below.

The intermediary $f_c$ values estimated for BD-14 1319 and IRAS 06475-0735 are in agreement
with the optically thick disc surrounded by a flared thin region proposed for Group I. 

HD 141549 shows the lowest value of $f_c$ that is consistent with the flat disc structure suggested for Group II. 

Among the four stars, IRAS 07394-1953 is the only one associated with nebulosity, as verified in the 
IRAS-ISIS images (see Sect. \ref{sectquality}), and probably for this reason it has a rising SED at the 
far-infrared that could not be fitted by HABdisk.

\subsection{Degeneracy of information from SED modelling}
\label{degeneracysect}

Despite the good quality of fittings, our results also indicate that some degenerated 
information from SED modelling can occur, as verified in the comparison of different models.

First, the flared disc model was not able to provide a good SED fitting when the observed disc size was 
considered for HD 141569.
A bad performance of the GA method is not the case, since a good fitting was achieved in a second run of 
the code, but leading to unrealistic disc parameters. 

Considering that a flat disc would better explain the circumstellar structure of HD 141569, 
we reanalyse the results from GH02 model, which gave reasonable disc size and temperature. 
However, other unrealistic parameters were also derived, probably due to the adopted optically thick regime
that gives high disc luminosity (see Table \ref{tabapptt}). 
This is not compatible with the optically thin disc that is observed in scattered light at large 
distances from the central star.
The disc parameters determination from the SED fitting remains unsolved since the radial structure of this disc 
is much more complicated than it is possible to describe using the adopted simple models.

\subsection{Flared vs. flat geometry}
\label{flatflaredsect}

On the view of the good results found for IRAS 06475-0735 and BD-14 1319, we conclude that
{\it PDS} stars within the $-1 < \beta_1 < 0.7$ range have SEDs better fitted by HABdisk instead of the flat disc model. 

For objects with $\beta_1 > 0.7$ another component seems to be required in the circumstellar structure, 
such as a shell for example, to better reproduce the observed excess at longer wavelengths. 
In this case, our previous flat disc model, which includes a thin dust envelope in the system,
provided better results than HABdisk.

Even considering the degeneracy in the case of HD 141569, its low $f_c$ value lead us to
expect that objects showing $\beta_1 < -1$ seem to be better explained by flat disc geometry models. 

\subsection{Conclusions}
\label{conclussect}

On the light of the above remarks, we conclude that 
good estimations of the circumstellar luminosities may be provided for a large sample of 
objects, and a statistical analysis can be done by comparing $f_c$ with $\beta_1$ in order to verify 
a possible sequence of disc disappearance. 

We are aware that no inferences about disc evolution can be attended from the derived parameters. 
Considering the lack of detailed observations about disc parameters or spectral features 
related to grain composition, a parameters set must be carefully evaluated, even when derived from 
good SED fittings. 
Since these detailed observations are not available for the large majority of the 
{\it PDS} sample, the disc parameters must be constrained to realistic ranges. 

The difficulty to infer disc parameters for large samples arises from the interpretation of details on 
infrared spectral features and/or high angular resolution data, requiring individual analysis of the physics, 
the geometry and chemical composition. 
This task shall be accomplished with the use of several promising high-sensitive observational facilities 
that are under operation or will be in a near future.
In particular, we mention here the near-infrared coronagraphic imager ({\it NICI}) that is being used with 
{\it Gemini} telescope to search for exoplanets. 
One of the sub-products of this campaign certainly shall be the detailed imagery of protoplanetary discs. 
Using the GA method, we seek to select good targets among the {\it PDS} stars that are in the 
distance range to be imaged with {\it NICI}.

\section*{Acknowledgments}

We acknowledge an anonymous referee for very useful comments on the improvement of the paper presentation. 
We are also grateful to Kees Dullemond for his kindness and valuable suggestions. JGH thanks 
partial support from FAPESP (Proc. No. 2005/00397-1). AHJ thanks IAG/USP for the opportunity to develop this 
research during a post-doctoral stay. This work has made use of the {\it SIMBAD } and {\it VizieR} databases
operated at CDS, Strasbourg, France.

\appendix

\section[]{Genetic Algorithms}
\label{gaapp}

GAs are a family of computer models inspired on natural evolution. Their basis is to assume a potential 
solution for a specific problem, viewed as chromosome like structure, on which are applied genetic operators 
(mutation, crossover, adaptation and evolution). The use of this technique simplifies the formulation and solution 
of optimisation problems, and parallel simultaneous procedure approach is implicit in the method, 
providing evaluation of the viability of a parameter set as possible 
solution for complex problems \citep{Koza92,Koza94,Holland92}.

The implementation of a GA starts with a random generation of sets of chromosomes 
$\left[ {S_1,S_2,\ldots,S_n } \right]$, which are evaluated and associated to an adaptation probability, $\chi^2$, 
obtained by the evaluation function, which is in our case the DDN01 model. The $\chi^2_i$ values express how 
each individual are adapted, or how each solution are close to the best solution \citep{Bentley02}. 

Then, the judgement function determines the genetic operator to be applied to a solution, and its values can 
be copy: the individual remains the same in the next generation, crossover: the individual is 
elected to change a number of genes (parameters) with another individual, creating a new one, mutation: 
one of its genes shall be randomly changed, or termination: none of the genes may continue on next 
generations. The chosen action is expressed by the $\Phi_i$ variable, associated to each individual. 
The further step is to evolve the current generation ($k$) to the next ($k+1$), what is done through a procedure 
$\Gamma$ that considers the solutions and the genetic operators designed by $\Phi_i$. Formally
\begin{equation}
\left[S_1,S_2,. .,S_n \right]_{k+1} =\Gamma \left[(S_1,\Phi_1),( S_2,\Phi_2),. .,(S_n,\Phi_n) \right]_k 
\end{equation}

\begin{figure}
\centerline{\includegraphics[width=2.0in,height=2.7in,angle=-90]{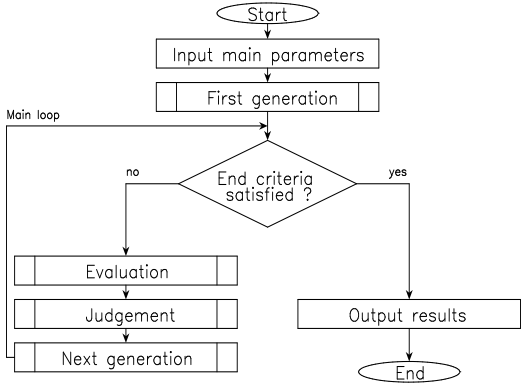}}
\caption{Schematic view of the main steps of a generic GA.}
\label{stepsGA}
\end{figure}

As soon as a new generation is ready, the evaluation function is reapplied, and the algorithm repeats the 
described actions until an end of loop condition is reached. The end condition can be based on the
number of iterations or quality (a low level for the $\chi^2_i$ values) (Figure \ref{stepsGA}). 

\section[]{Error bars estimative}
\label{errorapp}

Once an acceptable minimum has been found, one computes the inverse of the Hessian matrix 
\begin{equation}
[C] \equiv [\alpha]^{-1} 
\end{equation}
that has its components given by
\begin{equation}
\alpha _{ij} =\sum\limits_{k=1}^N {\left[ {\frac{\partial y(\lambda_k)}{\partial a_i }
\frac{\partial y(\lambda_k)}{\partial a_j }} \right]} 
\end{equation}
where $\partial y(\lambda_k) / \partial a_i$ means the partial derivative of the SED with respect to parameter 
$a_i$ at $\lambda = \lambda_k$, and $N$ is the number of observed data points. The components of the main 
diagonal of $C$ can be used to estimate the error bars on each parameter by \citep{Press95}: 
\begin{equation}
\sigma _i \simeq { \frac{C^{1/2}}{N} }
\end{equation}

\section[]{Flat disc model: review}
\label{ttapp}

\begin{table}
\centering
\caption{
The SED fitting parameters obtained by applying the flat disc model \protect\citep{GHH02} for the sample.}
\begin{tabular}{@{}lcccc@{}}
\hline
 & {\tiny HD 141569} & {\tiny BD-14 1319 } &{\tiny IRAS 06475} & {\tiny IRAS 07394} \\
\hline
$T_s$ (K) & 10110 & 10020 & 10170 & 10000 \\
$R_s$ (R$_\odot$) & 2.2 & 3.4 & 4.2 & 0.5 \\
$R_d$ (AU) & 383 & 35 & 387 & 657 \\
$\tau$ & 0.2 & 0.4 & 2.0 & 3.2 \\
$\theta$ ($^{\rm o}$) & 87& 54 & 0 & 2 \\
$R_e$ (AU) & 865 & 12000 & 566 & 730 \\
$gof$ & 0.019 & 0.010 & 0.019 & 0.043 \\
\hline
\end{tabular}
\label{tabapptt}
\end{table}

GH02 used a flat disc model assuming that the star and the disc are surrounded by a thin dust shell. 
The emission from the star and from the disc are attenuated by the opacity of the envelope. 
The flux is calculated by assuming blackbody emission and different temperature laws for the disc and the envelope. 

The model considers the following free parameters: 
radius $R_s$ of the star, radius $R_d$ and inclination $\theta$ of the 
disc, radius $R_e$ and total optical depth $\tau$ of the envelope. 
The stellar temperature $T_s$ is a fixed parameter, since it can be derived from the spectral type.
The absorption and the emission of radiation depend on $\tau$, 
where the extinction law is normalized to $\tau(1\mu {\rm m})=1$.
The stellar flux is expressed by
\begin{equation}
S_s \left( \lambda \right)=\frac{R_s^2 }{d^2}\pi B\left( {\lambda ,T_s } 
\right)e^{-\tau \left( \lambda \right)}
\end{equation}

A passive, optically thick and geometrically thin disc is assumed, 
for which the radiation is evaluated by adopting the temperature law given by \cite{adams86}:
\begin{equation}
T_d (r) =T_s \left( {\frac{2}{3\pi }} \right)^{1 \mathord{\left/ {\vphantom 
{1 4}} \right. \kern-\nulldelimiterspace} 4}\left( {\frac{r} {R_s }} 
\right)^{-3 \mathord{\left/ {\vphantom {3 4}} \right. 
\kern-\nulldelimiterspace} 4}
\end{equation}

The inner radius of the disc is defined by $R_h = R_s(T_g/T_s)^{-4/3}$, 
with adopted $T_g$ =1500 K as the destruction temperature of the grains, 
by considering the silicate sublimation temperature \citep{Chiang01}.
The disc contribution to the total emitted flux is given by 
\begin{equation}
S_d \left(\lambda \right)=\int_{R_h}^{R_d} \frac{2r \delta r}{d^2}\pi 
B\left[ \lambda ,T(r) \right] e^{-\tau \left( \lambda \right)}\cos \theta dr
\label{equSd}
\end{equation}

Following the radiative transfer model by \cite{epchtein90}, 
the emission from the dust envelope is calculated using the temperature law
\begin{equation}
T_e(r) =T_s \left( {\frac{2r }{R_s}} \right)^{-0.4}
\end{equation}

The optical depth depends on the density distribution: $\tau (r) \propto \rho \left( r \right)$,
where $\rho(r) \propto r^{-3/2}$.
The flux emitted by the envelope is obtained by 
\begin{equation}
S_e (\lambda )=\int_{R_d}^{R_e} \frac{{r^2 }{d^2}}\pi B\left[ \lambda ,T_e (r) \right]
e^{-\tau (r, \lambda )} \left( {1-e^{-\tau _{\rm e} \left(r, \lambda \right)}} \right) dr
\end{equation}
where $\tau _{\rm e}$ is the optical depth calculated from $r$ to $R_e$.

The total radiation emitted as a function of wavelength is expressed by
$S(\lambda)=S_s(\lambda)+S_d(\lambda)+S_e(\lambda)$.

Table \ref{tabapptt} gives the parameters obtained with this model. 
These results are used to compare the different morphologies assumed in the flat and flared models.

\label{lastpage}


\begin{thebibliography}{99}

\bibitem[\protect\citeauthoryear{Acke et al.}{2004}]{Acke04}Acke B., van den Ancker M. E., Dullemond C. P., van Boekel R., Waters L. B. F. M., 2004, A\&A, 422, 621
\bibitem[\protect\citeauthoryear{Adams \& Shu}{1986}]{adams86}Adams F.C., Shu F.H., 1986, ApJ, 308, 836 
\bibitem[\protect\citeauthoryear{Adams et al.}{1987}]{Adams87}Adams F. C., Lada C., Shu F. H., 1987, ApJ, 312, 788
\bibitem[\protect\citeauthoryear{Alencar et al.}{2003}]{Alencar03}Alencar S. H. P., Vaz L. P. R., Melo C. H. F., Dullmond C. P., Andersen J., Batalha C., Mathieu R. D., 2003, in L\'epine J. R.D., Gregorio-Hetem J., eds, proc. Open Issues in Local Star Formation, ASS 299, p.107
\bibitem[\protect\citeauthoryear{Andr\'{e} \& Montmerle}{1994}]{Andre94}Andr\'{e} P., Montmerle T., 1994, ApJ, 420, 837
\bibitem[\protect\citeauthoryear{Andr\'{e} et al.}{1993}]{Andre93}Andr\'{e} P., Ward-Thompson D., Barsony M., 1993, ApJ, 406, 122
\bibitem[\protect\citeauthoryear{Appenzeller \& Mundt}{1989}]{Appenz89}Appenzeller I., Mundt R., 1989, A\&ARv, 1, 291
\bibitem[\protect\citeauthoryear{Augereau et al.}{1999}]{Augereau99}Augereau J. C., Lagrange A. M., Mouillet D., M\'enard F., 1999, A\&A, 350, 51
\bibitem[\protect\citeauthoryear{Augereau \& Papaloizou}{2004}]{Augereau04}Augereau J. C., Papaloizou J. C. B., 2004, A\&A, 414, 1153
\bibitem[\protect\citeauthoryear{Bentley {\&} Corne}{2002}]{Bentley02}Bentley P. J., Corne D. W., 2002, Creative Evolutionary Systems. Morgan-Kaufmann, San Francisco
\bibitem[\protect\citeauthoryear{Bergin et al.}{2004}]{Bergin04}Bergin E., Calvet N., Sitko M. L., Abgrall H., D'Alessio P., et al., 2004, ApJ, 614, L133
\bibitem[\protect\citeauthoryear{Bessel et al.}{1998}]{Bess98}Bessel M. S., Castelli F., Plez B., 1998, A\&A, 333, 231
\bibitem[\protect\citeauthoryear{Boccaletti et al.}{2003}]{Boccaletti03}Boccaletti A., Augereau J. C., Marchis F., Hahn J., 2003, ApJ, 585, 494
\bibitem[\protect\citeauthoryear{Cardelli et al.}{1989}]{Cardelli89}Cardelli J. A., Cleyton G. C., Mathis J. S., 1989, ApJ, 345, 245
\bibitem[\protect\citeauthoryear{Carkner}{1998}]{Carkner98}Carkner, L. 1998, PhD Thesis, The Pennsylvania State University
\bibitem[\protect\citeauthoryear{Chiang {\&} Goldreich}{1997}]{Chiang97}Chiang E. I., Goldreich P., 1997, ApJ, 490, 368
\bibitem[\protect\citeauthoryear{Chiang et al.}{2001}]{Chiang01} Chiang E. I., Joung M. K., Creech-Eakman C. Q. I., Kesseler J. E., Blake G. A., van Dishoeck E. F., 2001, ApJ, 547, 1077 
\bibitem[\protect\citeauthoryear{Clampin et al.}{2003}]{Clampin03}Clampin M., Krist J. E., Ardila D. R., Golimowski D. A., et al., 2003, AJ, 126, 385
\bibitem[\protect\citeauthoryear{D'Alessio et al.}{2005}]{Dalessio05}D'Alessio P., Hartmann L., Calvet N., Franco-Hernández R., et al., 2005, ApJ, 621, 461
\bibitem[\protect\citeauthoryear{de Jager \& Nieuwenhuijzen}{1987}]{deJager87}de Jager C., Nieuwenhuijzen H., 1987, A\&A, 177, 217
\bibitem[\protect\citeauthoryear{Dominik et al.}{2003}]{DDWW03}Dominik C., Dullemond C. P., Waters L. B. F.M., Walch S., 2003, A\&A, 398, 607
\bibitem[\protect\citeauthoryear{Draine \& Lee}{1984}]{Draine84}Draine B. T., Lee H. M., 1984, ApJ, 285, 89
\bibitem[\protect\citeauthoryear{Dullemond \& Dominik}{2004}]{DD04}Dullemond C. P., Dominik C., 2004, A\&A, 417, 159
\bibitem[\protect\citeauthoryear{Dullemond et al.}{2001}]{DDN2001}Dullemond C. P., Dominik C., Natta A., 2001, ApJ, 560, 957
\bibitem[\protect\citeauthoryear{Dullemond}{2002}]{Dullemond02}Dullemond C. P., 2002, A\&A, 395, 853
\bibitem[\protect\citeauthoryear{Epchtein et al.}{1990}]{epchtein90} Epchtein N., Le Bertre T., L\'{e}pine J. R. D., 1990, A\&A, 227, 82
\bibitem[\protect\citeauthoryear{Feigelson \& Montmerle}{1999}]{Feigelson99}Feigelson, E.D., Montmerle, T. 1999, ARAA, 37, 363
\bibitem[\protect\citeauthoryear{Forrest et al.}{2004}]{Forrest04} Forrest W. J. 2004, ApJS, 154, 443
\bibitem[\protect\citeauthoryear{Geers et al.}{2006}]{Geers06}Geers V. C., Augerau J. C., Pontoppidan K. M., Dullemond C. P., et al., 2006, A\&A, 459, 545
\bibitem[\protect\citeauthoryear{Goldberg}{1989}]{Goldberg89}Goldberg D. E., 1989, Genetic Algorithms in Search, Optimization {\&} Machine Learning. Addison-Wesley Longman 
\bibitem[\protect\citeauthoryear{Gregorio-Hetem {\&} Hetem}{2002}]{GHH02}Gregorio-Hetem J., Hetem A. Jr., 2002, MNRAS, 336, 197
\bibitem[\protect\citeauthoryear{Gregorio-Hetem et al.}{1992}]{PDS1}Gregorio-Hetem J., L\'epine J. R. D., Quast G. R., Torres C. A. O., de la Reza R., 1992, AJ, 103, 549
\bibitem[\protect\citeauthoryear{Guimar\~aes et al.}{2006}]{Guimaraes06}Guimar\~aes M. M., Alencar S. H. P., Corradi W. J. B., Vieira S. L. A., 2006, A\&A, 457, 581 
\bibitem[\protect\citeauthoryear{Haisch et al.}{2001}]{Haisch01}Haisch K., Lada E., Lada C., 2001, ApJ, 553, L153 
\bibitem[\protect\citeauthoryear{Hern\'andez et al.}{2007}]{Hernandez07}Hern\'andez J., Hartmann L. W., Megeath T., Gutermuth R., et al., 2007, ApJ, 662, 1067
\bibitem[\protect\citeauthoryear{Hillenbrand et al.}{1992}]{Hillenbrand92}Hillenbrand L. A., Strom S. E., Vrba F. J., Keene J., 1992, ApJ, 397, 613
\bibitem[\protect\citeauthoryear{Holland}{1992}]{Holland92}Holland J. H., 1992, Adaptation in Natural and Artificial Systems. MIT Press 
\bibitem[\protect\citeauthoryear{Jonkheid et al.}{2006}]{Jonkheid06}Jonkheid B., Kamp I., Augerau J. C., van Dishoeck E. F., 2006, A\&A, 453, 163
\bibitem[\protect\citeauthoryear{Kenyon {\&} Hartmann}{1987}]{Kenyon87}Kenyon S. J., Hartmann L. W., 1987, ApJ, 323, 714
\bibitem[\protect\citeauthoryear{Koza}{1992}]{Koza92}Koza J. R., 1992, Genetic Programming: On the Programming of Computers by Means of Natural Selection. MIT Press 
\bibitem[\protect\citeauthoryear{Koza}{1994}]{Koza94}Koza J. R., 1994, Genetic Programming II: Automatic Discovery of Reusable Programs. MIT Press 
\bibitem[\protect\citeauthoryear{Lada \& Wilking}{1984}]{Lada84}Lada C. J., Wilking B. A., 1984, ApJ, 287, 610
\bibitem[\protect\citeauthoryear{Malfait et al.}{1998}]{Malfait98}Malfait K., Bogaert E., Waelkens C.,1998, A\&A, 331, 211
\bibitem[\protect\citeauthoryear{Meeus et al.}{2001}]{Meeus01}Meeus G., Waters L. B. F. M., Bouwman J., van den Ancker M. E., Waelkens C., Malfait K., 2001, A\&A, 365, 476
\bibitem[\protect\citeauthoryear{Meeus et al.}{2002}]{Meeus02}Meeus G., Bouwman J., Dominik C., Waters L. B. F. M., de Koter A., 2002, A\&A, 392, 1039
\bibitem[\protect\citeauthoryear{Mer\'\i n et al.}{2004}]{Merin04}Mer\'\i n B., Montesinos B., Eiroa C., Solano E., et al., 2004, A\&A, 419, 301
\bibitem[\protect\citeauthoryear{Miroshnichenko et al.}{1999}]{Mirosh99}Miroshnichenko A. S., Gray R. O., Vieira S. L. A., Kuratov K. S., Bergner, Y. K., 1999, A\&A, 347, 137
\bibitem[\protect\citeauthoryear{Mouillet et al.}{2001}]{Mouillet01}Mouillet D., Lagrange A. M., Augereau J. C., M\'enard F., 2001, A\&A, 372, L61
\bibitem[\protect\citeauthoryear{Press et al.}{1995}]{Press95} Press W. H., Teukolsky S. A., Vetterling W. T., Flannery B. P., 1995, Numerical Recipes in C - Second Edition. Cambridge University Press, New York
\bibitem[\protect\citeauthoryear{Sartori et al.}{2003}]{Sartori03}Sartori M. J., Gregorio-Hetem J., Hetem A. Jr., 2003, in L\'epine J. R.D., Gregorio-Hetem J., eds, proc. Open Issues in Local Star Formation, ASS 299, p.133
\bibitem[\protect\citeauthoryear{Savage \& Mathis}{1979}]{Savage79} Savage B. D., Mathis J. S., 1979, ARA\&A, 17, 73
\bibitem[\protect\citeauthoryear{Schultz \& Wiemer}{1975}]{Schultz06}Schultz G. V., Wiemer W., 1975 A\&A, 43, 133
\bibitem[\protect\citeauthoryear{Sicilia-Aguilar et al.}{2006}]{Sicilia06}Sicilia-Aguilar A., Hartmann L., Calvet N., Megeath S. T., et al., 2006, ApJ, 638, 897
\bibitem[\protect\citeauthoryear{Siess et al.}{2000}]{Siess00}Siess L., Dufour E., Forestine M., 2000, A\&A, 358, 593
\bibitem[\protect\citeauthoryear{Strom et al.}{1993}]{Strom93}Strom K. M., Strom S. E., Merrill K. M., 1993, ApJ, 412, 233
\bibitem[\protect\citeauthoryear{Sylvester et al.} {1996}]{Sylvester96}Sylvester R. J., Skinner C. J., Barlow M. J., Mannings V., 1996, MNRAS, 279, 915
\bibitem[\protect\citeauthoryear{Torres et al.}{1995}]{PDS2}Torres C. A. O., Quast, G. R., de la Reza R., Gregorio-Hetem J., L\'epine J. R. D., 1995, AJ, 109, 2146
\bibitem[\protect\citeauthoryear{Torres}{1998}]{Torres98}Torres C. A. O., 1998, Publica\c c\~ao Especial do Observat\'orio Nacional, No. 10/99
\bibitem[\protect\citeauthoryear{van den Ancker et al.}{2001}]{Ancker01}van den Ancker M. E., Meeus G., Cami J., Waters L. B. F. M., Waelkens C., 2001, A\&A, 369, 217
\bibitem[\protect\citeauthoryear{Vieira et al.}{2003}]{Vieira03}Vieira S. L. A., Corradi W. J. B., Alencar S. H. P., Mendes L. T. S., Torres C. A. O., Quast G. R., Guimar\~aes M. M., da Silva, L., 2003, AJ, 126, 2971
\bibitem[\protect\citeauthoryear{Weinberger et al.}{1999}]{Weinberger99}Weinberger A. J., Becklin E. E., Schneider G., Smith B. A., Lowrance P. J., Silverstone M. D., Zuckerman B., Terrile R. J., 1999, ApJ, 525, L53
\bibitem[\protect\citeauthoryear{Weinberger et al.}{2000}]{Weinberger00}Weinberger A. J., Rich R. M., Becklin E. E., Zuckerman B., Matthews K., 2000, ApJ, 544, 937
\bibitem[\protect\citeauthoryear{Weinberger et al.}{2004}]{Weinberger04}Weinberger A. J., Becklin E. E., Zuckerman B., 2004, A\&AS, 205, 1713
\bibitem[\protect\citeauthoryear{Wilking et al.}{1989}]{Wilking89}Wilking B. A., Lada C. J., Young E. T., 1989, ApJ, 340, 823
\bibitem[\protect\citeauthoryear{Yudin \& Evans}{1998}]{Yudin98}Yudin R. V., Evans A., 1998, A\&AS, 131, 401
\bibitem[\protect\citeauthoryear{Zuckerman et al.}{1995}]{Zuckerman95}Zuckerman B., Forveille T., Kastner J. H., 1995, Nature, 373, 494

\end{thebibliography}
\end{document}